\documentclass[sigconf]{acmart}
\AtBeginDocument{%
  \providecommand\BibTeX{{%
    \normalfont B\kern-0.5em{\scshape i\kern-0.25em b}\kern-0.8em\TeX}}}

\copyrightyear{2023}
\acmYear{2023}
\setcopyright{acmlicensed}\acmConference[WWW '23]{Proceedings of the ACM Web Conference 2023}{May 1--5, 2023}{Austin, TX, USA}
\acmBooktitle{Proceedings of the ACM Web Conference 2023 (WWW '23), May 1--5, 2023, Austin, TX, USA}
\acmPrice{15.00}
\acmDOI{10.1145/3543507.3583439}
\acmISBN{978-1-4503-9416-1/23/04}

\usepackage{booktabs}
\usepackage{tikz}
\usetikzlibrary{external}
\usepackage{ textcomp }
\usepackage[utf8]{inputenc}
\usepackage{enumitem}

\usepackage{epstopdf}
\usepackage{makecell}
\usepackage{threeparttable}
\usepackage{longtable}
\usepackage{multirow}
\usepackage{tabularx}
\usepackage{tabulary}
\usepackage{tablefootnote}
\usepackage{array}
\newcommand{\PreserveBackslash}[1]{\let\temp=\\#1\let\\=\temp}

\newcolumntype{C}[1]{>{\PreserveBackslash\centering}p{#1}}
\newcolumntype{R}[1]{>{\PreserveBackslash\raggedleft}p{#1}}
\newcolumntype{L}[1]{>{\PreserveBackslash\raggedright}p{#1}}
\usepackage{amssymb}
\usepackage{dsfont}
\usepackage{graphicx}
\usepackage{subfig}
\usepackage{caption}
\usepackage{color}
\usepackage{mathtools}
\usepackage{url}
\usepackage{hyperref}
\usepackage{bm}
\usepackage{algorithmic}
\usepackage[linesnumbered,ruled,vlined]{algorithm2e}

\SetCommentSty{mycommfont}
\begin{document}

\title{Multi-Behavior Recommendation with Cascading Graph Convolution Networks }
\author{ Zhiyong Cheng$^1$, Sai Han$^1$, Fan Liu$^2$, Lei Zhu$^{3,5}$, Zan Gao$^1$, Yuxin Peng$^{4,5}$}
\affiliation{\institution{$^1$Shandong Artificial Intelligence Institute, Qilu University of Technology (Shandong Academy of Sciences), \country{China}}}
\affiliation{\institution{$^2$ School of Computing, National University of Singapore, \country{Singapore}}}
\affiliation{\institution{$^3$School of
Information Science and Engineering, Shandong Normal University, \country{China}}}
\affiliation{\institution{$^4$Wangxuan Institute of Computer Technology, Peking University, \country{China}}}
\affiliation{\institution{$^5$Peng Cheng Laboratory, \country{China}}}
\email{{jason.zy.cheng,liufancs}@gmail.com,	hansai11@126.com}

\begin{abstract}
 Multi-behavior recommendation, which exploits auxiliary behaviors (e.g., \textit{click} and \textit{cart}) to help predict users’ potential interactions on the target behavior (e.g., \textit{buy}), is regarded as an effective way to alleviate the data sparsity or cold-start issues in recommendation. Multi-behaviors are often taken in certain orders  in real-world applications (e.g., \textit{click>cart>buy}). In a behavior chain, a latter behavior usually exhibits a stronger signal of user preference than the former one does. Most existing multi-behavior models fail to capture such dependencies in a behavior chain for embedding learning. In this work, we propose a novel multi-behavior recommendation model with cascading graph convolution networks (named MB-CGCN). In MB-CGCN, the embeddings learned from one behavior are used as the input features for the next behavior's embedding learning after a feature transformation operation. In this way, our model explicitly utilizes the behavior dependencies in embedding learning. Experiments on two benchmark datasets demonstrate the effectiveness of our model on exploiting multi-behavior data. It outperforms the best baseline by  33.7\% and 35.9\% on average over the two datasets in terms of Recall@10 and NDCG@10, respectively. 
\end{abstract}



\maketitle

\section{Introduction}
Recommendation, as an effective tool to deal with the information overload issue in the information era, has achieved a tremendous progress in the past decade. The model-based collaborative filtering, which learns user and item representations based on user-item interactions for recommendation, has become the mainstream recommendation technique since its success in the Netflix contest~\cite{MF}. With the rise of deep learning, CF methods have gained rapid development from shallow models~\cite{MF,BPRMF,cheng2018www} to deep models~\cite{NCF,cheng2022tois,NGCF,mao2021ultragcn}. The deep neural network (DNN) based models can learn better user/item representations and capture more complicated user-item interaction relations~\cite{LightGCN,LR-GCCF,DLsuvey,wu2022survey}. And the graph convolutional network (GCN) based ones exploit the high-order proximities over the user-item interaction graph to enhance the representation learning~\cite{NGCF,LightGCN,mao2021ultragcn,IMP-GCN}. In addition, the DNN- and GCN-based models are often equipped with advanced techniques, such as the attention mechanism~\cite{KGAT}, disentangled learning~\cite{DGCF}, contrastive learning~\cite{SimGCL,SGL}, setting up a new standard of recommendation. 

The aforementioned CF models are often built upon a single-type of behavior, which is directly related to the platform profit, such as the purchase behavior in e-commence platforms or the download behavior in an App platform. As such behaviors come with a real financial or time cost to users, single-behavior recommendation models will encounter serious data sparsity issue in real-world applications. Fortunately, users usually take other types of behaviors (i.e., \textit{click}, and \textit{cart}) to get more information about items to help them make the final decisions. For example, in an e-commerce platform, the \textit{click} behavior can bring users more information about the item (e.g., descriptions or reviews), and the \textit{cart} behavior is of the convenience to compare with other candidates for reaching a decision. Comparing to the purchase behaviors, the other types of behaviors have much richer interactions. And these types of behaviors or auxiliary behaviors can be exploited to help capture user preference and alleviate the data sparsity issue  in recommendation. 

Many multi-behavior recommendation models have been reported in literature~\cite{Tang2016multi,zhao2015www,NMTR,MBGCN,MBGMN}. A straightforward approach is to adapt the single-behavior recommendation methods to multi-behaviors, such as extending the standard matrix factorization to multiple matrices~\cite{Tang2016multi,zhao2015www} or designing new sampling strategies on the multiple behaviors based on BPR~\cite{ding2018ijcai,guo2017kbs,BPRH}. With the evidence of performance enhancement by using multi-behavior information, multi-behavior recommendation has drawn more attention in recent years. The recent advanced DNN and GCN techniques have been also applied in this task. For example, DIPN~\cite{DIPN} uses a hierarchical attention network to capture the inter- and intra-behavior relations; MATN~\cite{MATN} and KHGT~\cite{KHGT} apply the transformer network to model multi-behavior relations and learn user/item embeddings. The GCN-based models often construct a unified graph based on all types of interaction behaviors, and then perform GCN over the graph to learn the user and item embeddings with various strategies and techniques~\cite{KHGT,MBGCN,meng2022coarse,GHCF,MBGMN}. 

Despite the progress, those methods fail to exploit the behavior dependencies in a chain to directly facilitate the embedding learning in behaviors. In real-world applications, users often take behaviors in a certain order to discover more about an item to help make the final decision, such as \textit{click->cart->purchase}~\footnote{In real scenarios, users may jump one or two behaviors directly to the final behavior.}. Different behaviors reveal user preference to different extents or from different perspectives~\cite{MBGMN,CRGCN}. In a behavior chain, a latter behavior exhibits a stronger signal of user preference on the item than the former one does~\cite{chainRec}. Therefore, the preference information learned from a previous behavior can be used to facilitate the embedding learning in the next one in the behavior chain. Some methods also consider the dependencies or relations among multi-behaviors~\cite{MBGCN,DIPN,MATN,KHGT}, however, they learn the dependencies for weighing the contributions of other behaviors to the target behaviors in embedding aggregation. In addition, most existing methods treat each type of auxiliary behavior in the same way and have not considered the behavior order in modeling. Only limited methods consider the effects of behavior chain~\cite{NMTR,CRGCN}. NMTR~\cite{NMTR} models the cascading effects of multi-behaviors by injecting the prediction scores of a previous behavior into the next behavior’s score prediction. CRGCN~\cite{CRGCN} designs a cascading residual graph convolutional network to explicitly utilize the cascading behaviors in the embedding learning. Similar to many other multi-behavior recommendation models~\cite{DIPN,MGNN,MBGMN,NMTR,MATN,GHCF,meng2022coarse}, CRGCN also adopts multi-task learning in model training. Multi-task learning can acquire supervision signals from each type of behavior data for the embedding learning, but the learning process is not fully oriented to the target behavior prediction. 

In this paper, we propose a novel multi-behavior recommendation model with cascading graph convolution networks (named MB-CGCN). Specifically, our model consists of a sequence of GCN blocks, each of which corresponds to a behavior in the behavior chain. The LightGCN~\cite{LightGCN} is adopted for its efficiency and efficacy in the blocks to learn user and item embeddings from each behavior. The embeddings of users and items learned from a behavior are used as the input features to the next behavior's embedding learning.  Considering that directly using the output embeddings as the input embeddings of the next GCN may inject noise or misleading information to misguide the learning process, a feature transformation is designed to process the embeddings before the delivery. In this way, our model explicitly utilizes the behavior dependencies  in a chain to directly facilitate the embedding learning in latter behaviors. Finally, the embedding learned from different behaviors are aggregated for the final behavior prediction. Different from previous models, MB-CGCN does not adopt multi-task learning in optimization. It only uses the target behavior as supervision signals, making the embedding learning in all behaviors oriented to the ultimate goal. Extensive experiments have been conducted on two benchmark datasets to evaluate the effectiveness of  MB-CGCN. With the simple structure, MB-CGCN gains a remarkable improvement over the state-of-the-art multi-behavior recommendation methods, achieving 33.7\% and 35.9\% relative gains on average over the two datasets in terms of Recall@10 and NDCG@10, respectively. Further empirical studies are also performed to examine the validity of different designs in our model and analyze the effects of different multi-behavior numbers and orders. The main contributions are summarized as:
\vspace{-3pt}
\begin{itemize}[leftmargin=*]
    \item We highlight the importance of explicitly exploiting the behavior dependencies in a behavior chain  in multi-behavior recommendation and propose to distill the behavioral features of a former behavior to facilitate the embedding learning in the latter one.
    \item We propose a novel multi-behavior recommendation model MB-CGCN with a simple structure. It mainly consists of a sequence of GCN blocks, in which the embeddings learned in a previous GCN are processed by a feature transformation operation and then used as the input to the next one. 
    \item We comprehensively evaluate the effectiveness of MB-CGCN on two real-word datasets. Experiment results show that our model can significantly improve the recommendation performance with a large margin.  We release the our code for reproducibility\footnote{https://github.com/SS-00-SS/MBCGCN}. 
\end{itemize}

\section{Related work}

Multi-behavior recommendation refers to the exploitation of multiple behaviors in user-item interactions for recommendation~\cite{guo2017kbs,NMTR,MBGCN}. Owing to its effectiveness in alleviating the data sparsity issue and enhancing recommendation performance, it has drawn an increasing attention in recent years~\cite{BPRMF,NMTR,MBGCN,GNMR,meng2022coarse}. 

The early multi-behavior recommendation methods were mainly developed based on traditional recommendation techniques. A direct approach is to extend the traditional matrix factorization~{MF} technique operating on single matrix to multiple matrices~\cite{RLCMF,Tang2016multi}. For example, Ajit et al.~\cite{RLCMF} directly extended the matrix factorization model to factorize multiple matrices simultaneously with sharing item-side embeddings. This model was further extended to perform matrix factorization of multiple behaviors by sharing user or item embeddings~\cite{Tang2016multi,zhao2015www}. Another line of research treated the multiple behaviors as auxiliary behavioral data and designed new sampling strategies to enrich the training samples.  Loni et al.~\cite{Loni} proposed to assign different preference levels to multiple behaviors and extended the standard BPR~\cite{BPRMF} with a new negative sampling strategy for negative item sampling from different behaviors. Ding et al.~\cite{ding2018ijcai} further extended this idea by designing an improved negative sampler to better exploit the multiple behavioral data. Guo et al.~\cite{guo2017kbs} utilized the item-item similarity to generate samples from multiple auxiliary behaviors. Qiu et al.~\cite{BPRH} designed an adaptive sampler in BPR to balance the correlation among behaviors. 

With the great success of deep learning techniques in recommendation, researchers have also attempted to developed deep neural network (DNN) or graph convolutional network (GCN) based multi-behavior recommendation models in recent years. In DNN models, a common approach is to first learn user and item embeddings from each behavior via the designed network, and then aggregate the embeddings learned from different behaviors for the target behavior prediction~\cite{DIPN,MATN,KHGT}. The differences lie in the designed network and used attention mechanism. For example, DIPN~\cite{DIPN} applies a hierarchical attention network to model the relationships between different behaviors in embedding learning aggregation. MATN~\cite{MATN} uses a transformer-based network to encode the multi-behavior relations and a memory attention network to generate the embeddings for each behavior, which are then aggregated for target behavior prediction. Besides the user-item interactions, KHGT~\cite{KHGT} also considers the item-item relations and adopts a transformer network to learn user/item embeddings, which are aggregated with behavior-specific contributions to make the final prediction. Different from those methods that aggregate the embeddings from different behaviors for target behavior prediction, NMTR~\cite{NMTR} adopts a multi-task learning framework to leverage all behaviors of users as prediction targets and reuse the prediction score of the previous behavior to the score prediction of the next behavior. 

For GCN models, a general paradigm is to construct a unified user-item graph based on all behaviors, and then perform GCN operations on the graph to learn user embeddings~\cite{GHCF,MBGCN,MBGMN,meng2022coarse}. GHCF~\cite{GHCF} learns the embedding over the graph via GCN and then simply uses the aggregated representations to predict each behavior individually. MBGCN~\cite{MBGCN} learns the behavior contributions over the unified user-item graph and models behavior semantics over item-item co-interacted graphs. The final prediction is an aggregation of the predicted scores obtained by behavior contributions and behavior semantics. MGNN~\cite{MGNN} leverages the multiplex (multiple-layer) network to learn both shared user and item embeddings and unique embeddings of each behavior. MBGMN~\cite{MBGMN} uses a meta graph neural network to model the diverse multi-behavior patterns and capture the behavior heterogeneity and diversity in the unified graph. Meng et al.~\cite{meng2022coarse} considered the diversified interests of users behind multiple behaviors and proposed a knowledge-enhanced multi-interest learning method (CKML) to learn users’ multiple interests of users by assigning each behavior among interests. HMGGR~\cite{HMGGR} conducts graph contrastive learning among the constructed hyper meta-graphs to adaptively learn the complex dependencies among different behaviors for embedding learning. 

Besides chainRec~\cite{chainRec} and NMTR~\cite{NMTR}, the previous methods have not considered the order information in the modeling. And none of the aforementioned models have explicitly exploited the dependency relations of multi-behaviors into the embedding learning. More recently, Yan et al.~\cite{CRGCN} proposed a CRGCN model to address this limitation with a cascading GCN structure by using multi-task learning. However, due to the use of a residual design, it can only uses a single-layer GCN for auxiliary behaviors. In contrast, our model uses a feature transformation to distill features from previous behaviors and can avoid the problem caused by residual connections. Besides, we did not use the multi-task learning in optimization. 

\section{Proposed Model}

\begin{figure*}[htbp]
  \centering
  \includegraphics[width=0.9\textwidth]{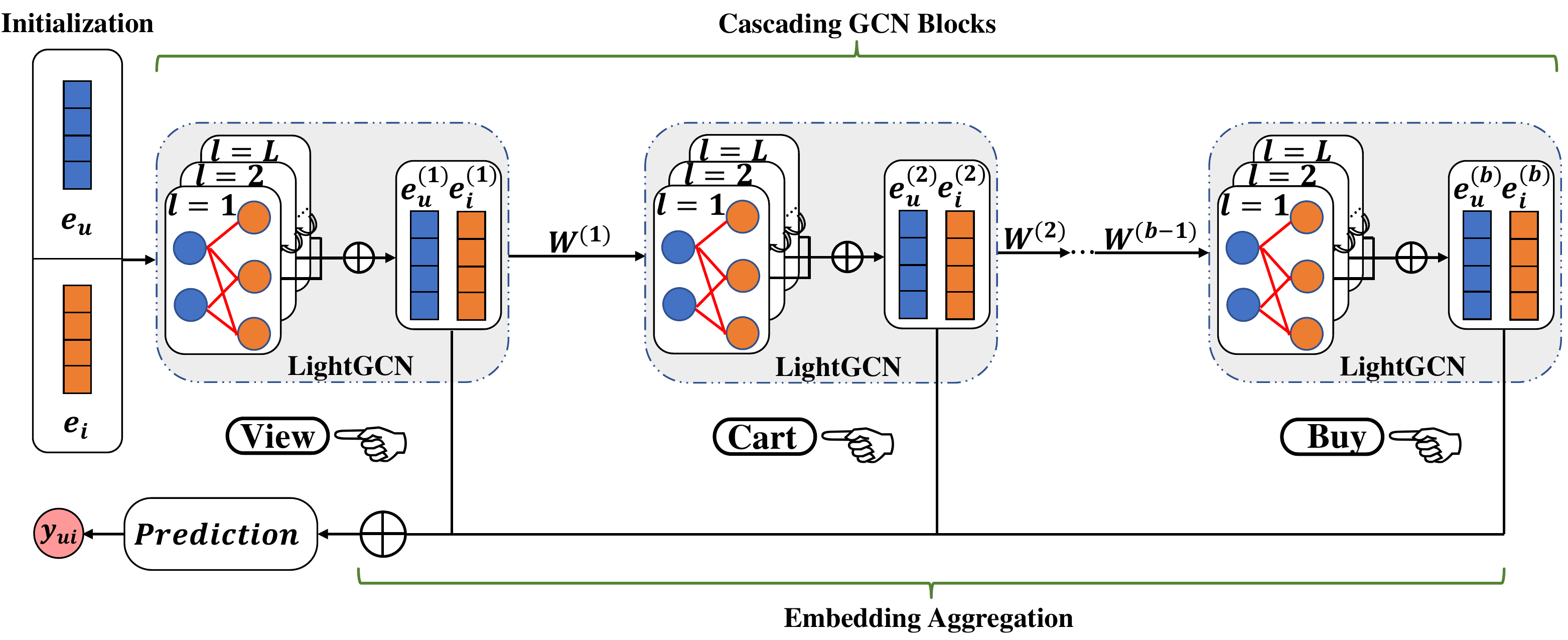}
  \caption{Overview of our MB-CGCN model.}
   \label{fig:overview}
\end{figure*}

\subsection{Problem Formulation}
Traditional recommendation methods are generally designed towards the single type of user-item interaction (i.e., a target behavior) that directly relevant to the platform's profit (e.g., \textit {buy}). They often face serious data sparsity or cold-start issue in real-world applications, because this behavior will bring a real financial cost to users. Before taking actions of this behavior, users will interact with items by other manners (e.g., \textit{click} and \textit{cart}) to get more information to help them make decisions. These auxiliary behaviors also contain rich information about user preferences and can be used to alleviate the data sparsity and cold-start issues. In this work, our goal is to design a recommendation model for the target behavior by exploiting the auxiliary behaviors.

Let $\mathcal{U}$ and $\mathcal{I}$ be the user set and item set with $M$ and $N$ users and items, respectively. We use $\{\bm{Y}^1,\bm{Y}^2,\cdots, \bm{Y}^{B}\}$ to denote the multi-behavior interaction matrices sorted in a behavior order. $B$ is the number of behavior types. $\bm{Y}^{b}$ is the interaction matrix of the $b$-th behavior and $\bm{Y}^{B}$ is the target behavior. All the interaction matrices are binary, which means that each entry in the matrices has a value 1 or 0, defined as:
\vspace{-3pt}
\begin{equation}
{y^b_{u,i}}=
 \begin{cases}
	1,& \text{If $u$ has interacted with $i$ under behavior $b$;}\\
	  0,& \text{otherwise.}
\end{cases}
\vspace{-3pt}
\end{equation}

The task of multi-behavior recommendation is formulated as:

\noindent \textbf{Input}: The interaction data of $B$ types of behaviors $\{\bm{Y}^1,\bm{Y}^2,\cdots, \bm{Y}^{B}\}$, for a user set $\mathcal{U}$ and an item set $\mathcal{I}$.

\noindent \textbf{Output}: A recommendation model to predict the probability  that a user $u$ will interact with an item $i$ under the $B$-th behavior, i.e., target behavior.

\subsection{Model Description}
\subsubsection{Overview}
Before delving into the details of our model, we would like to make a first impression on the global view. Motivated by the observations that different behaviors exhibit user preferences from different perspectives~\cite{meng2022coarse} or to different extents~\cite{chainRec,CRGCN}, we would like to explicitly utilize the dependencies among behavior chains for embedding learning in different behaviors. Figure~\ref{fig:overview} shows the overall structure of our MB-CGCN model, which consists of three components: 1) \textbf{Embedding initialization}, which initializes user and item embeddings for the subsequent learning based on the behaviors in a defined order. 2) \textbf{Cascading GCN blocks}, in which LightGCN is adopted to learn the user and item embeddings for each behaviors. More specifically, the embeddings learned from a previous behavior will be delivered to facilitate the next behavior's embedding learning via a feature transformation. 3) \textbf{Embedding aggregation}, which aggregates the embeddings learned from each behavior for the target behavior prediction.

\subsubsection{Embedding Initialization}
Following existing GCN-based recommendation models~\cite{LightGCN,mao2021ultragcn,MBGCN,CRGCN}, we initialize the embedding vectors of a user $u\in\mathcal{U}$ and an item $i\in\mathcal{I}$ as $\bm{e_u^0}\in \mathbb{R}^d$ and $\bm{e_i^0}\in \mathbb{R}^d$, respectively. $d$ denotes the embedding size. We use $\bm{P} \in \mathbb{R}^{M \times d}$ and $\bm{Q} \in \mathbb{R}^{N \times d}$ to respectively denote the embedding matrix for users and items. Each user or item is described by an unique ID, which is represented by an one-hot vector. Let $\bm{ID}^{U}$ and $\bm{ID}^{I}$ be the one-hot vector matrix for all users and items, the embedding of a user $u_m$ and item $i_n$ are initialized as:
\begin{equation}
  \label{eq:init}
  \begin{aligned}
    \bm{e}_{u_m}^{0} = \bm{P} \cdot \bm{ID}_m^{U}, \quad
    \bm{e}_{i_n}^{0} = \bm{Q} \cdot \bm{ID}_n^{I},
  \end{aligned}
\end{equation}
where $\bm{ID}_m^{U}$ and $\bm{ID}_n^{I}$ are one-hot vector for user $u_m$ and item $i_n$, respectively. Notice that the initialized embeddings are used as the input feature of users and items in the LightGCN of the first behavior, as shown in Figure~\ref{fig:overview}.

\subsubsection{Cascading GCN blocks}
The Cascading GCN blocks are the core component of our model. This component mainly consists of a LightGCN chain, in which each LightGCN~\cite{LightGCN} responses to one type of behaviors to learn the user and item embeddings for this behavior. Along with the chain, the embedding learned from a previous LightGCN is used as the input features of users and items in the next LightGCN after a \textit{feature transformation} operation. The basic idea is to use the cascading LightGCN to extract features from different behaviors and also exploit the dependencies in the behavior chain to help learn the latter behaviors' features. In the next, we briefly recap the core idea of LightGCN and introduce the feature transformation in our model.

\textbf{LightGCN Brief.}
LightGCN is an effective and popular GCN-based model designed for single-behavior recommendation. It removes the transformation matrix and nonlinear activation from the vanilla GCN. This simplification has proven to be efficient and effective in recommendation. In this work, we adopt LightGCN as the backbone model to learn user and item embeddings from each behavior for its efficiency and efficacy. Note other single-behavior GCN-based recommendation models can be also applied, such as UltraGCN~\cite{mao2021ultragcn} and SimGCL~\cite{SimGCL}. 

The core of GCN-based model is to recursively integrate the embedding information from neighboring nodes and update the embedding of ego nodes. Given the input embedding of a user $\bm{e}_u^b$ and an item $\bm{e}_i^b$ for the $b$-th behavior, LightGCN leverages the user-item interaction graph to propagate embeddings as:
\begin{equation}
{\bm{e^{(b,l+1)}_{u}}} ={\sum_{i\epsilon \mathcal{N}_u}  \frac{1}{\sqrt{\left\lvert \mathcal{N}_u\right\rvert} \sqrt{\left\lvert \mathcal{N}_i\right\rvert } } \bm{e^{(b,l)}_{i}}},
\end{equation}

\begin{equation}
{\bm{e^{(b,l+1)}_{i}}} ={\sum_{u\epsilon \mathcal{N}_i}  \frac{1}{\sqrt{\left\lvert \mathcal{N}_i\right\rvert} \sqrt{\left\lvert \mathcal{N}_u\right\rvert } } \bm{e^{(b,l)}_{u}}},
\end{equation}
where $\bm{e^{(b,l)}_{u}}$ and $\bm{e^{(b,l)}_{i}}$ respectively denote the updated embeddings of user $u$ and item $i$ under behavior $b$ after $l$ layers' propagation. $\mathcal{N}_u$ denotes the set of items that are interacted by user $u$, and $\mathcal{N}_i$ denotes the set of users that interact with item $i$. After $L$ layers' propagation, LightGCN obtains $L+1$ embeddings to describe a user $\{e_{u}^{(b,0)},e_{u}^{(b,1)},\cdots, e_{u}^{(b,L)}\}$ and an item $\{e_{i}^{(b,0)},e_{i}^{(b,1)},\cdots, e_{i}^{(b,L)}\}$. To obtain the final user and item embeddings based on the $b$-th behavior, we simply aggregate these embeddings as follows:
\begin{equation}
\bm{e_{u}^{(b)}} = \sum_{l = 0}^{L} \bm{e^{(b,l)}}_{u}, \quad \bm{e_{i}^{(b)}} = \sum_{l = 0}^{L} \bm{e^{(b,l)}_{i}}. 
\end{equation}
The learned embeddings from $b$-th behavior will fed into the LightGCN of the next behavior as the input embeddings after a \textit{feature transformation} operation. 

\textbf{Feature Transformation.}
All the different types of behaviors in interactions reveal users' preferences more or less. In a behavior chain, a latter behavior often exhibits stronger signal or more accurate user preference than a former behavior does~\cite{chainRec}. Therefore, the embeddings learned from a former behavior can be used as good initializations for  the next behavior's embedding learning, which is the underlying intuition of our cascading GCN structure. However, the direct use of a former behavior's features as initialized embeddings can be regarded as a refinement of the embeddings by using latter behaviors, which may lose the diverse information conveyed by different behaviors. On the other hand, the noisy information in a former behavior may negatively impact the learning process of latter behaviors seriously. With this consideration, we introduce a feature transformation design in MB-CGCN to process the learned embeddings  before the delivery. Let $\bm{W^{b}}$ be the transformation matrix for the $b$-th behavior to the $(b+1)$-th behavior, the transformation is performed as:

\begin{equation}
\bm{e_{u}^{(b+1,0)}} = \bm{W_{u}^b} \bm{e_{u}^{(b)}} , \quad \bm{e_{i}^{(b+1,0)}} = \bm{W_{i}^b} \bm{e_{i}^{(b)}},
\end{equation}
where $\bm{W_{u}^b}$ and $\bm{W_{i}^b}$ respectively denote the transformation vector for user $u$ and item $i$. $\bm{e_{u}^{(b+1,0)}}$ and $\bm{e_{i}^{(b+1,0)}}$ denote the initial embeddings of the user $u$ and item $i$ in the $(b+1)$-th behavior. Despite its simplicity, the feature transformation can effectively distill useful feature to facilitate the next behavior's embedding learning as demonstrated in our experiments. 

\subsubsection{Embedding Aggregation} \label{sec:disc}
To well exploit different behaviors, we aggregate the embeddings learned from all behaviors for prediction. The main focus of this work is to study the potential of exploiting the dependency structure of multi-behaviors in a certain order for recommendation. To keep the structure simple, we simply use a linear combination to aggregate the features learned from different behaviors, which is
\begin{equation}
\bm{e_{u}} = \sum_{b = 1}^{B} \bm{e^{(b)}_{u}}, \quad \bm{e_{i}} = \sum_{b = 1}^{B} \bm{e^{(b)}_{i}} . 
\end{equation}

Finally, the model prediction is defined as an inner product of the user and item embeddings:
\begin{equation}
\hat{y} _{ui} = \bm{e}^T_u \bm{e}_i,
\end{equation}
which is used as a ranking score for the target behavior recommendation.

\subsection{Model Training}
Similar to other rank-oriented recommendation work, the pairwise learning strategy is adopted for model optimization~\cite{LightGCN,IMP-GCN,MBGCN}. In implementation, we use the standard BPR loss~\cite{BPRMF}, which assumes a higher score to an observed item than that to an unobserved one. The objective function is formulated as:
\begin{equation}
	\mathcal{L} = \sum_{(u,i,j) \in O} -ln\sigma(y_{ui}-y_{uj})+\lambda \left\lVert \Theta \right\rVert^2,
\end{equation}
where $O=\{(u,i,j)|(u,i) \in \mathcal{R}^{+}, (u,j) \in \mathcal{R}^{-}\}$ is defined as positive and negative sample pairs, and $\mathcal{R}^{+}$ ($\mathcal{R}^{-}$) denotes the sample that has been observed (unobserved) in the target behavior. $\sigma(\cdot)$ denotes the sigmoid function.  $\Theta$ denotes all trainable parameters. $L_2$ regularization is adopted to prevent over-fitting and $\lambda$ is a coefficient to control the $L_2$ regularization.


\subsection{Discussion} \label{sec:discuss}
We notice that CRGCN~\cite{CRGCN} also exploits a cascading GCN structure to exploit the behavior dependencies in the embedding learning of  behaviors. As far as we concerned, MB-CGCN is fundamentally different from CRGCN on \emph{how to deliver the embedding from one behavior to the next behavior}. CRGCN delicately designs a residual connection to preserve previous behavioral features as initialized embeddings of the next behavior's network. In this way, it learns user and item embeddings  by gradually refining them through all the behaviors in the chain. The learned embeddings from the last behavior are directly used for prediction. Consequently, the quality of embeddings learned from earlier behaviors (e.g., \textit{click}) exacts a great impact on the final performance. Because the earlier behaviors are not deterministic and often noisy, using higher-order propagation for embedding learning in such behaviors is inevitably bring more noise into embeddings. This also explains why CRGCN uses only one-layer propagation in auxiliary behaviors. In contrast, we adopt a feature transformation to distill useful information for next behavior's embedding learning. This helps our model avoid the above problems, and thus our model can enjoy the benefits brought by high-order propagation. 


\section{Experiment}


\subsection{Experiment Settings}
\subsubsection{Dataset}
Two datasets are adopted for evaluation:

\begin{itemize}[leftmargin=*]
\item \textbf{Beibei}: This dataset was collected from Beibei\footnote{https://www.beibei.com/}, which is the largest e-commerce platform for baby products in China. It contains 21,716 users and 7,977 items with three types of user-item behaviors, including \textit{view}, \textit{adding-to-cart} or \textit{cart} for short, and \textit{buy}. Notice that on the Beibei platform, users have to follow a strict order to make purchase, which is \textit{view>cart>buy}.
\item \textbf{Tmall}: This dataset was collected from Tmall\footnote{https://www.tmall.com/}, one of the largest e-commerce platforms in China. It contains 15,449 users and 11,953 items. We also use the three types of behaviors in this dataset as those in Beibei for experiments.
\end{itemize}

For both datasets, we followed the previous studies to remove the duplicates by keeping the earliest one~\cite{NMTR,MBGCN}. 
The statistical information of the two datasets used in our experiments is summarized in Table~\ref{tab:freq}.

\begin{table}
  \caption{Statistics of the datasets used in our experiments.}
  \label{tab:freq}
  \begin{tabular}{cccccc}
     \toprule
      Dataset & User\# & Item\# & Buy\# & Cart\# & View\# \\
      \midrule
      Beibei & 21,716 & 7,997  & 304,576 & 642,622 & 2,412,586 \\
      Tmall  & 15,449 & 11,953 & 104,329 & 195,476 & 873,954 \\
      \bottomrule
\end{tabular}
\end{table}

\begin{table*}[htb]
  \caption{Overall performance comparison.  Improv. denotes the relative improvements over the best baseline.}
  \label{tab:result}
  \resizebox{\linewidth}{!}{
    \setlength{\tabcolsep}{3mm}{
      \begin{tabular}{c|l|ccc|ccccc|c|c}
        \hline
        \multirow{2}*{Dataset} & \multirow{2}{*}{Metric} & \multicolumn{3}{c|}{\textbf{Single behavior Methods}} & \multicolumn{6}{c|}{\textbf{Multi behavior Methods}} & \multirow{2}*{Improv.}                                                                                              \\
        \cline{3-11}         &                         & \textbf{MF-BPR}           & \textbf{NeuMF}            & \textbf{LightGCN}  & \textbf{RGCN}    & \textbf{GNMR}    & \textbf{NMTR}    & \textbf{MBGCN}               & \textbf{CRGCN}  & \textbf{MB-CGCN}   &                   
        \\
        \hline
        \multirow{6}*{Beibei}  & Recall@10           & 0.0191               & 0.0232             & 0.0391         & 0.0363 & 0.0413   & 0.0429    & \underline{0.0470}  & 0.0459              & \textbf{0.0579}  & 23.2\%  \\
                               & NDCG@10             & 0.0049               & 0.0135             & 0.0209         & 0.0188 & 0.0221   & 0.0198    & 0.0259              & \underline{0.0324}  & \textbf{0.0381}  & 17.6\%   \\
                               & Recall@20           & 0.0531               & 0.0736             & 0.0717         & 0.0684 & 0.0729   & 0.0776    & 0.0792              & \underline{0.0891}  & \textbf{0.0972}  & 9.1\%  \\
                               & NDCG@20             & 0.0239               & 0.0290             & 0.0270         & 0.0274 & 0.0279   & 0.0296    & 0.0330              & \underline{0.0348}  & \textbf{0.0404}  & 16.1\%  \\
                               & Recall@50           & 0.1014               & 0.1402             & 0.1347         & 0.1309 & 0.1391   & 0.1453    & 0.1493              & \underline{0.1694}  & \textbf{0.1924}  & 13.6\%  \\
                               & NDCG@50             & 0.0330               & 0.0405             & 0.0366         & 0.0371 & 0.0374   & 0.0399    & 0.0447              & \underline{0.0487}  & \textbf{0.0572}  & 17.5\%  \\
                               
        \hline
        \multirow{6}*{Tmall}   & Recall@10           & 0.0076               & 0.0236             & 0.0411         & 0.0215 & 0.0368   & 0.0282    & 0.0509   & \underline{0.0855}  & \textbf{0.1233} & 44.2\%  \\
                               & NDCG@10             & 0.0036               & 0.0128             & 0.0240         & 0.0104 & 0.0216   & 0.0137    & 0.0294   & \underline{0.0439}  & \textbf{0.0677} & 54.2\%   \\
                               & Recall@20           & 0.0244               & 0.0311             & 0.0546         & 0.0326 & 0.0608   & 0.0642    & 0.0691   & \underline{0.1369} &\textbf{0.2007}  & 46.6\%  \\
                               & NDCG@20             & 0.0155               & 0.0152             & 0.0266         & 0.0125 & 0.0263   & 0.0303    & 0.0350   & \underline{0.0676} &\textbf{0.0880}  & 30.2\%  \\
                               & Recall@50           & 0.0393               & 0.0494             & 0.0874         & 0.0411 & 0.0971   & 0.1034    & 0.1117   & \underline{0.2325} &\textbf{0.3322}  & 42.9\%  \\
                               & NDCG@50             & 0.0197               & 0.0193             & 0.0338         & 0.0160 & 0.0336   & 0.0383    & 0.0455   & \underline{0.0866} &\textbf{0.1134}  & 30.9\%  \\
        \hline
      \end{tabular}
    }
  }
  \vspace{-3pt}
\end{table*}

\subsubsection{Evaluation Protocols}
We adopt the widely used leave-one-out strategy for evaluation~\cite{NMTR,MBGCN,MBGMN}. For each user, the last interacted item and all the items which she has not interacted with comprise the test set; the second last interacted item of each user is used to construct the validation set for hyper-parameter tuning; the remainder positive items are used in training.  In the evaluation stage, all the items in the test set are ranked according to the predicted scores by recommendation models.  In our experiments, two standard metrics for top-n recommendation \textit{Recall@K} and \textit{NDCG@K} are used to measure performance.

\subsubsection{Baselines}
To demonstrate the effectiveness of our MB-CGCN, we compare it with several state-of-the-art methods, which can be classified into two categories: single-behavior models and mujlti-behavior models

\noindent \textbf{Single-behavior Models:}
\begin{itemize}[leftmargin=*]
\item \textbf{MF-BPR}\cite{BPRMF}: This method has shown competitive performance on the top-n recommendation task, and is commonly used as a baseline to evaluate the efficacy of new models. BPR is a widely used optimization strategy with the assumption that the positive items should score higher than negative ones.

\item \textbf{NeuMF}\cite{NCF}: It is a representative neural CF models, which uses GMF and MLP simultaneously to capture the non-linear interactions between users and items.

\item \textbf{LightGCN}\cite{LightGCN}: This model is a state-of-the-art GCN-based recommendation model, which exploits the high-order connectives in the user-item bipartite graph for recommendation. In particular, it removes the feature transformation and non-linear activation components in vanilla GCN to simplify the model structure and achieves a significant performance improvement.
\end{itemize}

\noindent  \textbf{Multi-behavior Models:}
\begin{itemize}[leftmargin=*]

\item \textbf{NMTR}\cite{NMTR}: This model develops
a cascading neural network method to model the multi-behavior data for recommendation. It sequentially passes the interaction score of the current behavior to the next and uses multi-task learning for jointly optimization.

\item \textbf{RGCN}\cite{RGCN}: This model differentiates the relations between nodes via edge types in the graph and designs different propagation layers for different relations. This model can adapt to multi-behavior recommendation

\item \textbf{GNMR}\cite{GNMR}: This GNN-based approach attempts to explore the dependencies among multi-behaviors via recursive embedding propagation on the unified multi-behavior interaction graph. It designs a relation aggregation network to  model the interaction heterogeneity.

\item \textbf{MBGCN}\cite{MBGCN}: It is a state-of-the-art GCN-based multi-behavior recommendation model. This method considers the different contributions of multi-behaviors to the target behavior. It learns the behavior contributions by using GCN on the unified multi-behavior graph and exploits the item-item graph to capture the behavior semantics.

\item \textbf{CRGCN}\cite{CRGCN}: This is a most recently proposed model. It  adopts a cascading GCN structure to model multi-behavior data. The behavioral features learned  from a behavior is delivered to the next behavior with a residual design. This method also adopts the multi-task learning in optimization. 

\end{itemize}

\subsubsection{Hyper-parameter Settings.} In implementation, we adopt the \textit{Adam} optimizer for optimization. The embedding size and the batch size of all adopted methods are set to 64 and 1024, respectively. The learning rate is tuned in the range of \{1e{-2}, 1e{-3}, 1e{-4}\}. In addition, we initialize model parameters (behavior features transformation matrices) with the Xavier initializer. The early stopping strategy is also adopted.   The number of GCN layers for each behaviors is tuned in \{1, 2, 3, 4\}. Without specification, for the behavior chain \textit{view>cart>buy}, we use the layers of $\{3,4,3\}$ and $\{3,4,2\}$ for the three behaviors on Beibei and Tmall, respectively. For other baselines, we mainly use their official open-source code and carefully tune the parameters to achieve the best performance for fair comparisons.
%


\subsection{Overall Performance}
In this section, we report the performance comparisons between our MB-CGCN and all the baselines. The results on two datasets are shown in Table~\ref{tab:result}. The best results are highlighted in bold and the second best results are underlined. From the results, we can see that the multi-behavior models can achieve better performance than the single-behavior models, demonstrating the benefits of leveraging auxiliary behaviors (i.e., view and cart) for the target behavior (i.e., buy) prediction. Our MB-CGCN model achieves the best performance,  outperforming all baselines significantly in term of both metrics over two datasets. The improvement across different ranges of top $K$ ($K=\{10,20,50\}$) items over the best baseline can achieve 9.0\% and 16.1\% on Beibei, and 46.6\% and 30.1\% on Tmall for Recall@20 and NDCG@20 metrics, respectively. It is a remarkable improvement for recommendation accuracy, especially observing that the best baseline CRGCN has already made a big improvement over the second best baseline, which strongly demonstrates the effectiveness of our MB-CRGN model.


For the single-behavior models, NeuMF outperforms MF-BPR in most cases. MF-BPR uses the inner product as interaction function, which is incapable of modeling the complex relations between users and items. In contrast, NeuMF adopts multi-layers of neural networks to model the non-linear interactions, yielding a better performance. LightGCN achieves consistently better performance over MF-BPR and NeuMF. This demonstrates the advantages of the GCN models by exploiting the high-order neighbor's information over the user-item bipartite graph to learn user and item embeddings for recommendation.

For the multi-behavior recommendation models, RGCN models different behaviors individually and then aggregates the embeddings learned from each behavior for prediction without distinguishing their contributions to the target behavior. It does not perform well among the multi-behavior models. Both GNMR and MBGCN differentiate the behavior contributions before fusion, and they achieve a better performance than RGCN. Comparing to GNMR, MBGCN additionally exploits the item-item relations to capture the behavior semantics and gains further improvement over GNMR. NMTR considers the cascading effects of multi-behaviors in the model structure. It attempts to take the effects into the embedding learning process by passing the prediction scores of a previous behavior to the next one. It surpasses the GNMR model but does not perform as well as MBGCN. The better performance MBGCN attributes to its use of GCN and additional consideration of item-item relations in the modeling. CRGCN moves a step further over NMTR by directly take the cascading effects of multi-behavior into the embedding learning process explicitly. It is achieved by passing the embeddings learned from the previous behavior to the next one for further refinement. In this way, the embedding learning process of CRGCN is actually a refinement of the embeddings through the behavior chain. CRGCN outperforms all other baselines by a large margin, especially on the Tmall datasets. 

MB-CGCN adopts the similar cascading GCN structure as CRGCN, and thus also enjoys the merits of explicitly exploiting the cascading effects in embedding learning. Instead of adopting the residual design in CRGCN to preserve behavior features for delivery, MB-CGCN adopts a feature transformation operation between two GCN blocks to distill effective features from a previous behavior to the next one. In addition, MB-CGCN does not use the multi-task learning in optimization and only employ the signals of target behavior to guide the learning process. The big performance improvement of MB-CGCN over CRGCN demonstrates the effectiveness of our design.



\begin{table}
  \caption{Effects of the feature transformation in MB-CGCN. The reported performance is computed based on the top 20 results.($w/o.\ ft$ and $w.\ ft$ denote MB-CGCN with and without the feature transformation, respectively. }

    \begin{tabular}{r|cc|cc} 
        \hline
        \textbf{\multirow{2}{*}{Method}} & \multicolumn{2}{c|}{\textbf{Beibei}} & \multicolumn{2}{c}{\textbf{Tmall}} \\
        \cline{2-5}
                & \textbf{Recall} & \textbf{NDCG} & \textbf{Recall} & \textbf{NDCG}\\
		\hline
		\textbf{w/o. ft} & 0.0892 & 0.0382 & 0.1994 & 0.0825\\
		\textbf{w. ft}   & \textbf{0.0972} & \textbf{0.0404} & \textbf{0.2007} & \textbf{0.0880}\\
    \hline
    \end{tabular}
\label{tab:ft}
\end{table}

\subsection{Ablation Study}
\subsubsection{Effect of feature transformation}
In order to evaluate the validity of the feature transformation in our model, we conducted an ablation study to compare our model with and without the feature transformation module (denoted as \textbf{w/o. ft} and \textbf{w. ft} in Table 3). 
 The default order of behaviors used in our experiments are: \textit{view>cart>buy}\footnote{There could be other orders, such as \textit{view>buy}, \textit{cart>buy}, and \textit{cart>view>buy}, which are studied in Section~\ref{sec:behaviororder}.}, which also the required order in Beibei. After removing the feature transformation from MB-CGCN,  the embeddings learned from the first behavior (i.e., \textit{view}) are directly used as the initialized embeddings in the next behavior (i.e., \textit{cart}) for embedding learning; and the same for the embedding learning of the last behavior (i.e., \textit{buy}).

Experimental results are shown in Table~\ref{tab:ft}. With feature transformation, MB-CGCN can gain an relative improvement  of 9.0\% and 5.8\% on Beibei, and 0.7\% and 6.7\% on Tmall for Recall@20 and NDCG@20\%, respectively. A latter behavior in the behavior chain usually shows a strong signal of user preferences on items. The results demonstrate the effectiveness of the feature transformation scheme on distilling useful information from an earlier behavior to help learn user and item embeddings in a latter behavior.



\subsubsection{Effects of feature aggregation}
 In MB-CGCN, we aggregate the embeddings learned from all behaviors for users and items for the target behavior prediction by a linear aggregation. To evaluate the utility of the feature aggregation,  we compare our model to two variants in experiments:
\begin{itemize}[leftmargin=*]
\item \textbf{w/o. agg.:} This variant removes the feature aggregation module in MB-CGCN. It means the embeddings learned from the last GCN block are directly used for the target prediction.
\item \textbf{w/o. concat.:} This variant replaces the aggregation with a concatenation operation. Specifically, the user and item embeddings learned from each behavior are concatenated together for the target prediction.  
\end{itemize}

From the results shown in Table~\ref{tab:agg}, it is clear that it is necessary to consider the embedding learned from all behaviors. As discussed, with the feature transformation, some features learned from auxiliary behaviors will be filtered when delivering the embeddings learned from the first behavior to the target behavior. And it will also encourage the model to learn different features from each behavior. Therefore, to well exploit the multi-behaviors, it is important for our model to consider all the behavioral features. For simplicity, we only compare the linear combination (\textbf{w. agg}) with the embedding concatenation (\textbf{w. concat.}) in experiments. Empirically, \textbf{w. agg} performs much better than \textbf{w. concat.}. More sophisticated fusion methods (e.g., attention network) can be also applied, which will be explored in future studies.



\begin{table}
  \caption{Effects of feature aggregation in MB-CGCN. The reported performance is computed based on the top 20 results.}
    \begin{tabular}{r|cc|cc}
        \hline
        \multirow{2}{*}{\textbf{Method}} & \multicolumn{2}{c|}{\textbf{Beibei}} & \multicolumn{2}{c}{\textbf{Tmall}} \\
        \cline{2-5}
        & \textbf{Recall} & \textbf{NDCG} & \textbf{Recall} & \textbf{NDCG}\\
		\hline
		\textbf{w/o. agg.}  & 0.0556 & 0.0140 & 0.0698 & 0.0291\\
		\textbf{w.  concat.} & 0.0758 & 0.0282 & 0.1648 & 0.0688\\
		\textbf{w. agg.}    & \textbf{0.0972} & \textbf{0.0404} & \textbf{0.2007} & \textbf{0.0880}\\
    \hline
    \end{tabular}
    \label{tab:agg}
\end{table}

\subsection{Impact of multi-behaviors}

\subsubsection{Behavior number.}
To study the effects of auxiliary behaviors on the performance of our model, we perform experiments on the two datasets with one behavior (i.e., \textit{buy}), two behaviors (e.g., \textit{cart>buy}), and three behaviors (e.g., \textit{view>cart>buy}). Experimental results are reported in Table~\ref{tab:number}. Notice that MB-CGCN with one behavior is simplified to the LightGCN. Apparently, with more types of behaviors, our model performs better. Interestingly, with the use of \textit{cart} data, the performance is improved by a large margin; with the further consideration of \textit{view} data, the improvement becomes smaller. The reasons might be twofold. First, the \textit{cart} data have already enrich the user-item interactions to a large extent, it becomes harder to improve the performance. Besides, the \textit{cart} data can better uncover user preference than the view data. Notice that a \textit{view} behavior does not necessarily means that a user is interested in an item. 


\begin{table}
  \caption{Effects of behavior number in MB-CGCN. The reported performance is computed based on the top 20 results.}

    \begin{tabular}{r|cc|cc}
        \hline
        \multirow{2}{*}{\textbf{Method}} & \multicolumn{2}{c|}{\textbf{Beibei}} & \multicolumn{2}{c}{\textbf{Tmall}} \\
        \cline{2-5}
            & \textbf{Recall} & \textbf{NDCG} & \textbf{Recall} & \textbf{NDCG}\\
		\hline
		\textbf{buy}           & 0.0717   & 0.0270 & 0.0546   & 0.0266\\
		\textbf{cart>buy}      & 0.0930   & 0.0389 & 0.1956   & 0.0851\\
		\textbf{view>cart>buy} & \textbf{0.0972}   & \textbf{0.0404} & \textbf{0.2007}   & \textbf{0.0880}\\
    \hline
    \end{tabular}
\label{tab:number}
\end{table}



    

\subsubsection{Behavior order.} \label{sec:behaviororder}
\begin{figure}[h]
  \centering
  \includegraphics[width=\linewidth]{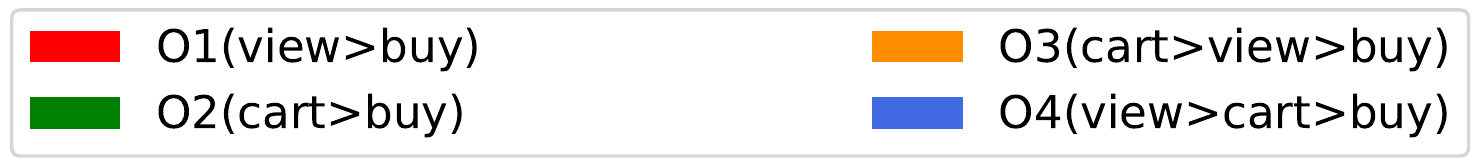}
  \includegraphics[width=\linewidth]{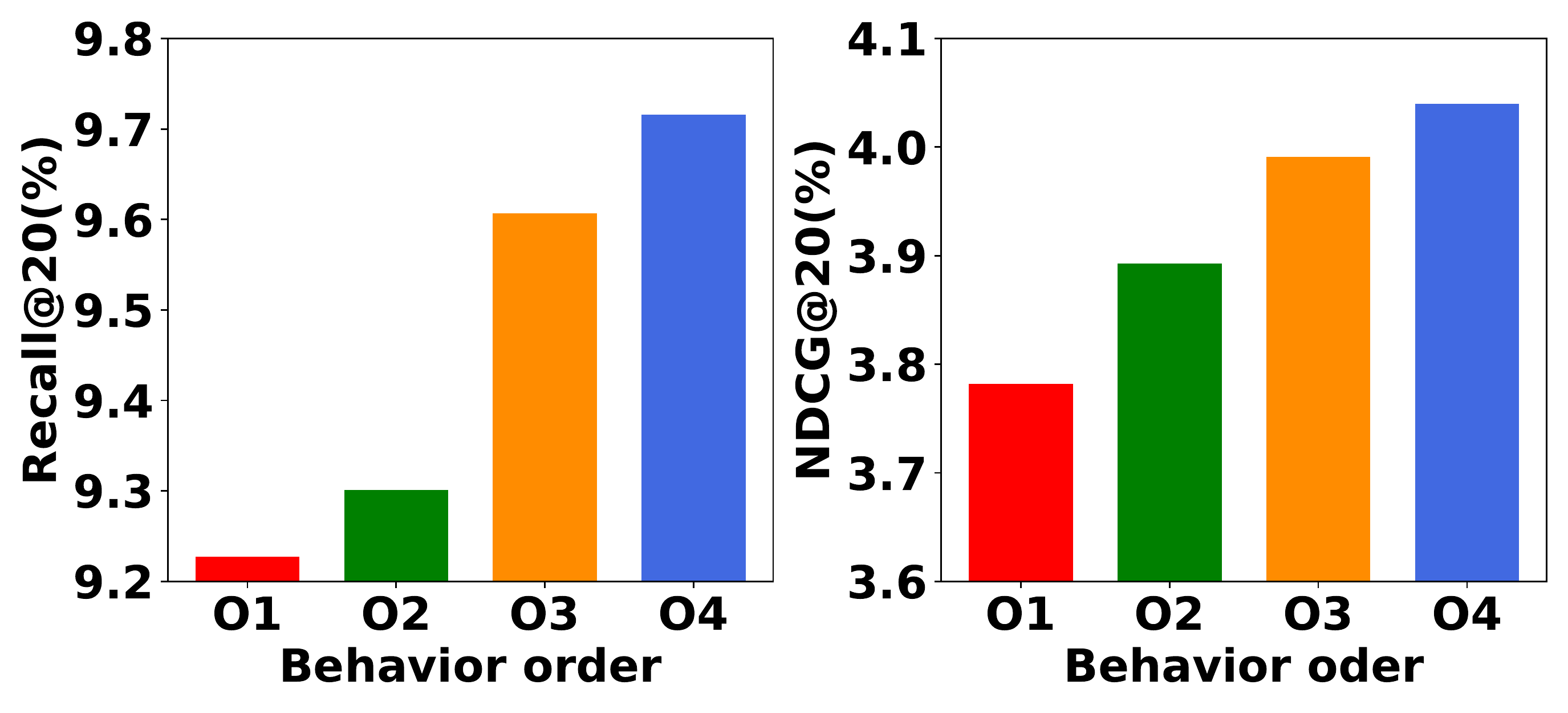}
  \caption{Effects of behavior order on Beibei.}
  \label{fig:bbei}
\end{figure}

\begin{figure}[h]
  \centering
  \includegraphics[width=\linewidth]{label.pdf}
  \includegraphics[width=\linewidth]{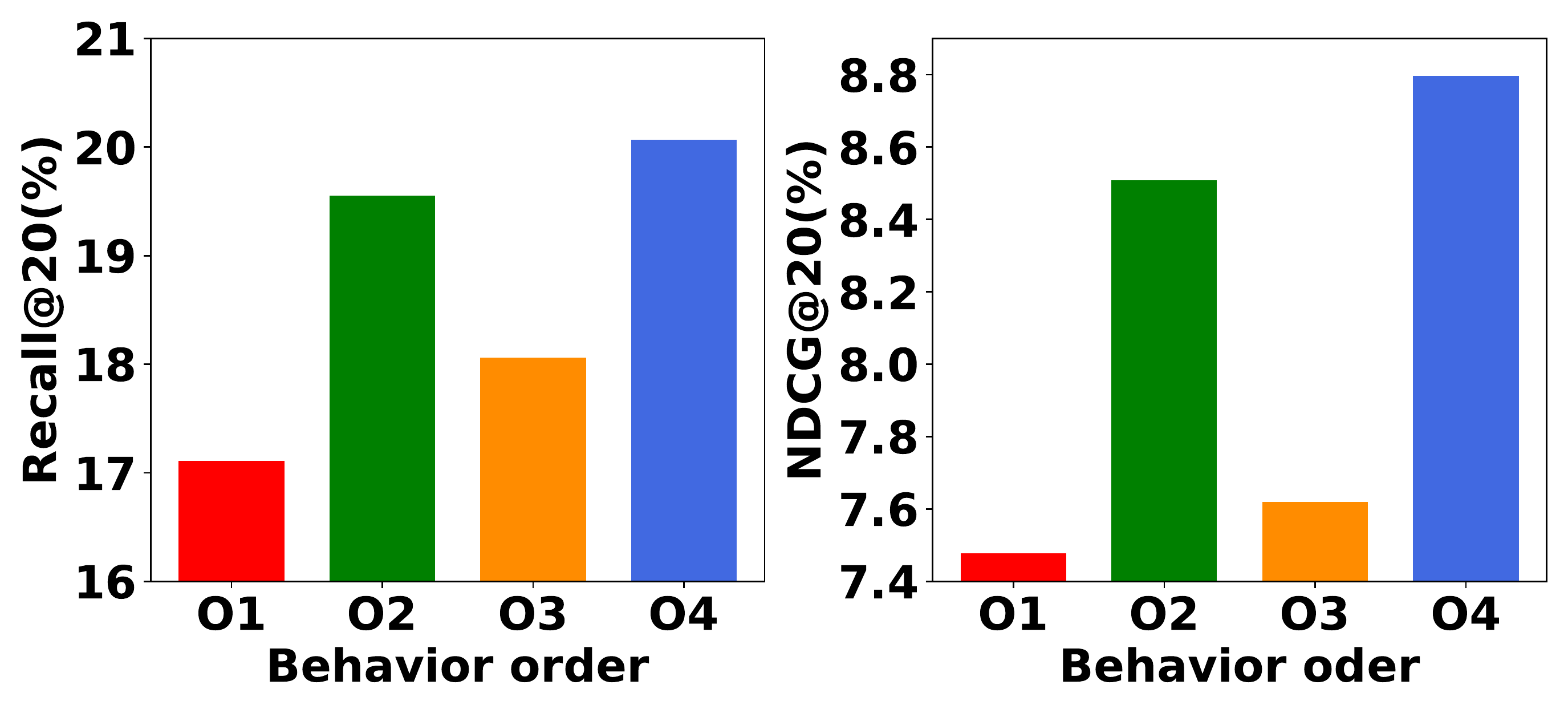}
  \caption{Effects of behavior order on Tmall.}
  \label{fig:btmall}
\end{figure}

We study the effects of behavior order on our model by four behavior orders on the two datasets: O1(\textit{cart>buy}), O2(\textit{view>cart}), O3(\textit{cart>view>buy}), and O4(\textit{view>cart>buy}). The performance of different behavior orders on two datasets are shown in Figure~\ref{fig:bbei} and Figure~\ref{fig:btmall}. 
Because both O3 and O4 have the same behaviors with different orders, we first compare their performance. The performance of O4 is better than O3 over both datasets, demonstrating the importance of modeling the multi-behaviors in a correct order, which should be consistent with the common behavior orders in real-world applications. Though the interactions of \textit{view} is much richer over that of \textit{cart} on both datasets, the performance of O2 is much better than that of O1. This is because the \textit{cart} behavior reveals more accurate information about user
preferences than the \textit{view} information as discussed above. It is surprising that our model with O3 performs worse than it with O2 on Tmall. Remind that the embedding learned from a previous behavior will directly affect the embedding learning in the next behavior due to the cascading design of our model. A reasonable behavior order, in which a latter behavior should reveal user preference more accurate than its previous behavior, can make the embeddings of the target behavior be gradually learned step by step through the behavior chain. By using the behavior order in O3, the noisy information in the \textit{view} data will negatively affect the embedding learning in the target behavior, whose features play an important role in the final prediction.  This further demonstrates the importance of the behavior order in the cascading modeling of multi-behaviors.


\vspace{-6pt}

\subsection{Effects of layer number}
 
In the this study, we tune the layer number by setting it the same for all the behaviors.\footnote{Notice that we also performed the studied by separately tuning the number of layers for the LightGCN of each behavior, which does not reported here due to the space limitation. We deem the reported experiments here can already provide a global view on the effects of layer number.} Table~\ref{tab:layer} reports the performance of our model with one, two, three layers on both datasets. MB-CGCN performs better over both datasets with the increase of layer numbers. With more layers, LightGCN can exploit higher-order information in the user-item interaction graph for embedding learning, as demonstrated in previous work~\cite{LightGCN,IMP-GCN}. The better embedding learned from each behavior benefits our model for final prediction. 

It is worth mentioning that CRGCN, which uses the similar structure as our model, cannot benefit from the high-order GCN on the auxiliary behaviors~\cite{CRGCN}. This is because CRGCN uses a residual design to preserve the behavior features in embedding delivery. However, there could be misleading interactions in auxiliary behaviors (such as \textit{view}). For those behaviors, exploiting higher-order neighbors will bring more noise to the model and hurt the performance eventually. In our model, a feature transformation module is used to process the embedding for delivery. This enables our model to enjoy the benefits from high-order GCN operations.  
\begin{table}
  \caption{Effects of layer numbers by setting the same layer numbers to all behaviors.}


    \begin{tabular}{r|cc|cc}
        \hline
        \multirow{2}{*}{\textbf{Method}} & \multicolumn{2}{c|}{\textbf{Beibei}} & \multicolumn{2}{c}{\textbf{Tmall}} \\
        \cline{2-5}
        & \textbf{Recall} & \textbf{NDCG} & \textbf{Recall} & \textbf{NDCG}\\
		\hline
		\textbf{1-Layer}         & 0.0942   & 0.0328 & 0.1923   & 0.0864\\
		\textbf{2-Layer}         & 0.0954   & 0.0359 & 0.1933   & 0.0867\\
		\textbf{3-Layer}   & \textbf{0.0961}   & \textbf{0.0370} & \textbf{0.1967}   & \textbf{0.0869}\\
    \hline
    \end{tabular}
\label{tab:layer}
\vspace{-3pt}
\end{table}

\vspace{-3pt}
\section{conclusion}
In this work, we present a novel multi-behavior recommendation model named MB-CGCN, which adopts the cascading GCN blocks to explicitly leverage the multi-behaviors for embedding learning. In this model, the behavior features learned by LightGCN over a previous behavior is delivered to the next behavior in a chain after a feature transformation operation. The embedding learned from all behaviors are aggregated for the final prediction. Experiments on two real-world datasets show that MB-CGCN outperforms the state-of-the-art multi-behavior models with a substantial performance gain. Further ablation studies verify the effectiveness of the feature transformation and embedding aggregation in our model. We also study the impact of multi-behavior number and order on the final performance. For future work, we plan to conduct experiments on online systems with A/B testing to validate the utility of our model on the real-world applications.

\vspace{-3pt}
\section*{Acknowledgments}
This work was supported in part by the National Key R\&D Program of China under Grant 2021YFF0901502; in part by the National Natural Science Foundation of China under Grants 61902223, 62272254, 62132001, and 61925201; in part by the Shandong Project towards the Integration of Education and Industry under Grants 2022PY009, and 2022PYI001.



\bibliographystyle{ACM-Reference-Format}
\bibliography{mbcgcn}


\begin{thebibliography}{37}


\ifx \showCODEN    \undefined \def \showCODEN     #1{\unskip}     \fi
\ifx \showDOI      \undefined \def \showDOI       #1{#1}\fi
\ifx \showISBNx    \undefined \def \showISBNx     #1{\unskip}     \fi
\ifx \showISBNxiii \undefined \def \showISBNxiii  #1{\unskip}     \fi
\ifx \showISSN     \undefined \def \showISSN      #1{\unskip}     \fi
\ifx \showLCCN     \undefined \def \showLCCN      #1{\unskip}     \fi
\ifx \shownote     \undefined \def \shownote      #1{#1}          \fi
\ifx \showarticletitle \undefined \def \showarticletitle #1{#1}   \fi
\ifx \showURL      \undefined \def \showURL       {\relax}        \fi
\providecommand\bibfield[2]{#2}
\providecommand\bibinfo[2]{#2}
\providecommand\natexlab[1]{#1}
\providecommand\showeprint[2][]{arXiv:#2}

\bibitem[Chen et~al\mbox{.}(2021)]%
        {GHCF}
\bibfield{author}{\bibinfo{person}{Chong Chen}, \bibinfo{person}{Weizhi Ma},
  \bibinfo{person}{Min Zhang}, \bibinfo{person}{Zhaowei Wang},
  \bibinfo{person}{Xiuqiang He}, \bibinfo{person}{Chenyang Wang},
  \bibinfo{person}{Yiqun Liu}, {and} \bibinfo{person}{Shaoping Ma}.}
  \bibinfo{year}{2021}\natexlab{}.
\newblock \showarticletitle{Graph Heterogeneous Multi-Relational
  Recommendation}. In \bibinfo{booktitle}{\emph{Proceedings of the Thirty-Fifth
  {AAAI} Conference on Artificial Intelligence}}. \bibinfo{publisher}{{AAAI}
  Press}, \bibinfo{pages}{3958--3966}.
\newblock


\bibitem[Chen et~al\mbox{.}(2020)]%
        {LR-GCCF}
\bibfield{author}{\bibinfo{person}{Lei Chen}, \bibinfo{person}{Le Wu},
  \bibinfo{person}{Richang Hong}, \bibinfo{person}{Kun Zhang}, {and}
  \bibinfo{person}{Meng Wang}.} \bibinfo{year}{2020}\natexlab{}.
\newblock \showarticletitle{Revisiting Graph Based Collaborative Filtering: {A}
  Linear Residual Graph Convolutional Network Approach}. In
  \bibinfo{booktitle}{\emph{The Thirty-Fourth {AAAI} Conference on Artificial
  Intelligence}}. \bibinfo{publisher}{{AAAI} Press}, \bibinfo{pages}{27--34}.
\newblock


\bibitem[Cheng et~al\mbox{.}(2018)]%
        {cheng2018www}
\bibfield{author}{\bibinfo{person}{Zhiyong Cheng}, \bibinfo{person}{Ying Ding},
  \bibinfo{person}{Lei Zhu}, {and} \bibinfo{person}{Mohan~S. Kankanhalli}.}
  \bibinfo{year}{2018}\natexlab{}.
\newblock \showarticletitle{Aspect-Aware Latent Factor Model: Rating Prediction
  with Ratings and Reviews}. In \bibinfo{booktitle}{\emph{Proceedings of the
  2018 World Wide Web Conference on World Wide Web}}.
  \bibinfo{publisher}{{ACM}}, \bibinfo{pages}{639--648}.
\newblock


\bibitem[Cheng et~al\mbox{.}(2022)]%
        {cheng2022tois}
\bibfield{author}{\bibinfo{person}{Zhiyong Cheng}, \bibinfo{person}{Fan Liu},
  \bibinfo{person}{Shenghan Mei}, \bibinfo{person}{Yangyang Guo},
  \bibinfo{person}{Lei Zhu}, {and} \bibinfo{person}{Liqiang Nie}.}
  \bibinfo{year}{2022}\natexlab{}.
\newblock \showarticletitle{Feature-Level Attentive {ICF} for Recommendation}.
\newblock \bibinfo{journal}{\emph{{ACM} Trans. Inf. Syst.}}
  \bibinfo{volume}{40}, \bibinfo{number}{4} (\bibinfo{year}{2022}),
  \bibinfo{pages}{75:1--75:24}.
\newblock


\bibitem[Ding et~al\mbox{.}(2018)]%
        {ding2018ijcai}
\bibfield{author}{\bibinfo{person}{Jingtao Ding}, \bibinfo{person}{Guanghui
  Yu}, \bibinfo{person}{Xiangnan He}, \bibinfo{person}{Yuhan Quan},
  \bibinfo{person}{Yong Li}, \bibinfo{person}{Tat{-}Seng Chua},
  \bibinfo{person}{Depeng Jin}, {and} \bibinfo{person}{Jiajie Yu}.}
  \bibinfo{year}{2018}\natexlab{}.
\newblock \showarticletitle{Improving Implicit Recommender Systems with View
  Data}. In \bibinfo{booktitle}{\emph{Proceedings of the Twenty-Seventh {IJCAI}
  International Joint Conference on Artificial Intelligence}}.
  \bibinfo{publisher}{ijcai.org}, \bibinfo{pages}{3343--3349}.
\newblock


\bibitem[Gao et~al\mbox{.}(2021)]%
        {NMTR}
\bibfield{author}{\bibinfo{person}{Chen Gao}, \bibinfo{person}{Xiangnan He},
  \bibinfo{person}{Dahua Gan}, \bibinfo{person}{Xiangning Chen},
  \bibinfo{person}{Fuli Feng}, \bibinfo{person}{Yong Li},
  \bibinfo{person}{Tat{-}Seng Chua}, \bibinfo{person}{Lina Yao},
  \bibinfo{person}{Yang Song}, {and} \bibinfo{person}{Depeng Jin}.}
  \bibinfo{year}{2021}\natexlab{}.
\newblock \showarticletitle{Learning to Recommend With Multiple Cascading
  Behaviors}.
\newblock \bibinfo{journal}{\emph{{IEEE} Trans. Knowl. Data Eng.}}
  \bibinfo{volume}{33}, \bibinfo{number}{6} (\bibinfo{year}{2021}),
  \bibinfo{pages}{2588--2601}.
\newblock


\bibitem[Guo et~al\mbox{.}(2017)]%
        {guo2017kbs}
\bibfield{author}{\bibinfo{person}{Guibing Guo}, \bibinfo{person}{Huihuai Qiu},
  \bibinfo{person}{Zhenhua Tan}, \bibinfo{person}{Yuan Liu},
  \bibinfo{person}{Jing Ma}, {and} \bibinfo{person}{Xingwei Wang}.}
  \bibinfo{year}{2017}\natexlab{}.
\newblock \showarticletitle{Resolving data sparsity by multi-type auxiliary
  implicit feedback for recommender systems}.
\newblock \bibinfo{journal}{\emph{Knowl. Based Syst.}}  \bibinfo{volume}{138}
  (\bibinfo{year}{2017}), \bibinfo{pages}{202--207}.
\newblock


\bibitem[Guo et~al\mbox{.}(2019)]%
        {DIPN}
\bibfield{author}{\bibinfo{person}{Long Guo}, \bibinfo{person}{Lifeng Hua},
  \bibinfo{person}{Rongfei Jia}, \bibinfo{person}{Binqiang Zhao},
  \bibinfo{person}{Xiaobo Wang}, {and} \bibinfo{person}{Bin Cui}.}
  \bibinfo{year}{2019}\natexlab{}.
\newblock \showarticletitle{Buying or Browsing?: Predicting Real-time
  Purchasing Intent using Attention-based Deep Network with Multiple Behavior}.
  In \bibinfo{booktitle}{\emph{Proceedings of the 25th {ACM} {SIGKDD}
  International Conference on Knowledge Discovery {\&} Data Mining}}.
  \bibinfo{publisher}{{ACM}}, \bibinfo{pages}{1984--1992}.
\newblock


\bibitem[He et~al\mbox{.}(2020)]%
        {LightGCN}
\bibfield{author}{\bibinfo{person}{Xiangnan He}, \bibinfo{person}{Kuan Deng},
  \bibinfo{person}{Xiang Wang}, \bibinfo{person}{Yan Li},
  \bibinfo{person}{Yong{-}Dong Zhang}, {and} \bibinfo{person}{Meng Wang}.}
  \bibinfo{year}{2020}\natexlab{}.
\newblock \showarticletitle{LightGCN: Simplifying and Powering Graph
  Convolution Network for Recommendation}. In
  \bibinfo{booktitle}{\emph{Proceedings of the 43rd International {ACM} {SIGIR}
  conference on research and development in Information Retrieval}}.
  \bibinfo{publisher}{{ACM}}, \bibinfo{pages}{639--648}.
\newblock


\bibitem[He et~al\mbox{.}(2017)]%
        {NCF}
\bibfield{author}{\bibinfo{person}{Xiangnan He}, \bibinfo{person}{Lizi Liao},
  \bibinfo{person}{Hanwang Zhang}, \bibinfo{person}{Liqiang Nie},
  \bibinfo{person}{Xia Hu}, {and} \bibinfo{person}{Tat{-}Seng Chua}.}
  \bibinfo{year}{2017}\natexlab{}.
\newblock \showarticletitle{Neural Collaborative Filtering}. In
  \bibinfo{booktitle}{\emph{Proceedings of the 26th {WWW} International
  Conference on World Wide Web}}. \bibinfo{publisher}{{ACM}},
  \bibinfo{pages}{173--182}.
\newblock


\bibitem[Jin et~al\mbox{.}(2020)]%
        {MBGCN}
\bibfield{author}{\bibinfo{person}{Bowen Jin}, \bibinfo{person}{Chen Gao},
  \bibinfo{person}{Xiangnan He}, \bibinfo{person}{Depeng Jin}, {and}
  \bibinfo{person}{Yong Li}.} \bibinfo{year}{2020}\natexlab{}.
\newblock \showarticletitle{Multi-behavior Recommendation with Graph
  Convolutional Networks}. In \bibinfo{booktitle}{\emph{Proceedings of the 43rd
  International {ACM} {SIGIR} conference on research and development in
  Information Retrieval}}. \bibinfo{publisher}{{ACM}},
  \bibinfo{pages}{659--668}.
\newblock


\bibitem[Koren et~al\mbox{.}(2009)]%
        {MF}
\bibfield{author}{\bibinfo{person}{Yehuda Koren}, \bibinfo{person}{Robert~M.
  Bell}, {and} \bibinfo{person}{Chris Volinsky}.}
  \bibinfo{year}{2009}\natexlab{}.
\newblock \showarticletitle{Matrix Factorization Techniques for Recommender
  Systems}.
\newblock \bibinfo{journal}{\emph{Computer}} \bibinfo{volume}{42},
  \bibinfo{number}{8} (\bibinfo{year}{2009}), \bibinfo{pages}{30--37}.
\newblock


\bibitem[Liu et~al\mbox{.}(2021)]%
        {IMP-GCN}
\bibfield{author}{\bibinfo{person}{Fan Liu}, \bibinfo{person}{Zhiyong Cheng},
  \bibinfo{person}{Lei Zhu}, \bibinfo{person}{Zan Gao}, {and}
  \bibinfo{person}{Liqiang Nie}.} \bibinfo{year}{2021}\natexlab{}.
\newblock \showarticletitle{Interest-aware Message-Passing {GCN} for
  Recommendation}. In \bibinfo{booktitle}{\emph{Proceedings of the The Web
  Conference 2021}}. \bibinfo{publisher}{{ACM} / {IW3C2}},
  \bibinfo{pages}{1296--1305}.
\newblock


\bibitem[Loni et~al\mbox{.}(2016)]%
        {Loni}
\bibfield{author}{\bibinfo{person}{Babak Loni}, \bibinfo{person}{Roberto
  Pagano}, \bibinfo{person}{Martha~A. Larson}, {and} \bibinfo{person}{Alan
  Hanjalic}.} \bibinfo{year}{2016}\natexlab{}.
\newblock \showarticletitle{Bayesian Personalized Ranking with Multi-Channel
  User Feedback}. In \bibinfo{booktitle}{\emph{Proceedings of the 10th {ACM}
  Conference on Recommender Systems}}. \bibinfo{publisher}{{ACM}},
  \bibinfo{pages}{361--364}.
\newblock


\bibitem[Mao et~al\mbox{.}(2021)]%
        {mao2021ultragcn}
\bibfield{author}{\bibinfo{person}{Kelong Mao}, \bibinfo{person}{Jieming Zhu},
  \bibinfo{person}{Xi Xiao}, \bibinfo{person}{Biao Lu},
  \bibinfo{person}{Zhaowei Wang}, {and} \bibinfo{person}{Xiuqiang He}.}
  \bibinfo{year}{2021}\natexlab{}.
\newblock \showarticletitle{UltraGCN: Ultra Simplification of Graph
  Convolutional Networks for Recommendation}. In
  \bibinfo{booktitle}{\emph{Proceedings of the 30th {ACM} International
  Conference on Information and Knowledge Management}}.
  \bibinfo{publisher}{{ACM}}, \bibinfo{pages}{1253--1262}.
\newblock


\bibitem[Meng et~al\mbox{.}(2022)]%
        {meng2022coarse}
\bibfield{author}{\bibinfo{person}{Chang Meng}, \bibinfo{person}{Ziqi Zhao},
  \bibinfo{person}{Wei Guo}, \bibinfo{person}{Yingxue Zhang},
  \bibinfo{person}{Haolun Wu}, \bibinfo{person}{Chen Gao},
  \bibinfo{person}{Dong Li}, \bibinfo{person}{Xiu Li}, {and}
  \bibinfo{person}{Ruiming Tang}.} \bibinfo{year}{2022}\natexlab{}.
\newblock \showarticletitle{Coarse-to-Fine Knowledge-Enhanced Multi-Interest
  Learning Framework for Multi-Behavior Recommendation}.
\newblock \bibinfo{journal}{\emph{CoRR}}  \bibinfo{volume}{abs/2208.01849}
  (\bibinfo{year}{2022}).
\newblock


\bibitem[Qiu et~al\mbox{.}(2018)]%
        {BPRH}
\bibfield{author}{\bibinfo{person}{Huihuai Qiu}, \bibinfo{person}{Yun Liu},
  \bibinfo{person}{Guibing Guo}, \bibinfo{person}{Zhu Sun},
  \bibinfo{person}{Jie Zhang}, {and} \bibinfo{person}{Hai~Thanh Nguyen}.}
  \bibinfo{year}{2018}\natexlab{}.
\newblock \showarticletitle{{BPRH:} Bayesian personalized ranking for
  heterogeneous implicit feedback}.
\newblock \bibinfo{journal}{\emph{Inf. Sci.}}  \bibinfo{volume}{453}
  (\bibinfo{year}{2018}), \bibinfo{pages}{80--98}.
\newblock


\bibitem[Rendle et~al\mbox{.}(2009)]%
        {BPRMF}
\bibfield{author}{\bibinfo{person}{Steffen Rendle}, \bibinfo{person}{Christoph
  Freudenthaler}, \bibinfo{person}{Zeno Gantner}, {and} \bibinfo{person}{Lars
  Schmidt{-}Thieme}.} \bibinfo{year}{2009}\natexlab{}.
\newblock \showarticletitle{{BPR:} Bayesian Personalized Ranking from Implicit
  Feedback}. In \bibinfo{booktitle}{\emph{Proceedings of the Twenty-Fifth {UAI}
  Conference on Uncertainty in Artificial Intelligence}}.
  \bibinfo{publisher}{{AUAI} Press}, \bibinfo{pages}{452--461}.
\newblock


\bibitem[Schlichtkrull et~al\mbox{.}(2018)]%
        {RGCN}
\bibfield{author}{\bibinfo{person}{Michael~Sejr Schlichtkrull},
  \bibinfo{person}{Thomas~N. Kipf}, \bibinfo{person}{Peter Bloem},
  \bibinfo{person}{Rianne van~den Berg}, \bibinfo{person}{Ivan Titov}, {and}
  \bibinfo{person}{Max Welling}.} \bibinfo{year}{2018}\natexlab{}.
\newblock \showarticletitle{Modeling Relational Data with Graph Convolutional
  Networks}. In \bibinfo{booktitle}{\emph{Proceedings of the 15th {ESWC}
  International Conference on Semantic Web}} \emph{(\bibinfo{series}{Lecture
  Notes in Computer Science}, Vol.~\bibinfo{volume}{10843})}.
  \bibinfo{publisher}{Springer}, \bibinfo{pages}{593--607}.
\newblock


\bibitem[Singh and Gordon(2008)]%
        {RLCMF}
\bibfield{author}{\bibinfo{person}{Ajit~Paul Singh} {and}
  \bibinfo{person}{Geoffrey~J. Gordon}.} \bibinfo{year}{2008}\natexlab{}.
\newblock \showarticletitle{Relational learning via collective matrix
  factorization}. In \bibinfo{booktitle}{\emph{Proceedings of the 14th {ACM}
  {SIGKDD} International Conference on Knowledge Discovery and Data Mining}}.
  \bibinfo{publisher}{{ACM}}, \bibinfo{pages}{650--658}.
\newblock


\bibitem[Tang et~al\mbox{.}(2016)]%
        {Tang2016multi}
\bibfield{author}{\bibinfo{person}{Liang Tang}, \bibinfo{person}{Bo Long},
  \bibinfo{person}{Bee{-}Chung Chen}, {and} \bibinfo{person}{Deepak Agarwal}.}
  \bibinfo{year}{2016}\natexlab{}.
\newblock \showarticletitle{An Empirical Study on Recommendation with Multiple
  Types of Feedback}. In \bibinfo{booktitle}{\emph{Proceedings of the 22nd
  {ACM} {SIGKDD} International Conference on Knowledge Discovery and Data
  Mining}}. \bibinfo{publisher}{{ACM}}, \bibinfo{pages}{283--292}.
\newblock


\bibitem[Wan and McAuley(2018)]%
        {chainRec}
\bibfield{author}{\bibinfo{person}{Mengting Wan} {and}
  \bibinfo{person}{Julian~J. McAuley}.} \bibinfo{year}{2018}\natexlab{}.
\newblock \showarticletitle{Item recommendation on monotonic behavior chains}.
  In \bibinfo{booktitle}{\emph{Proceedings of the 12th {ACM} Conference on
  Recommender Systems}}. \bibinfo{publisher}{{ACM}}, \bibinfo{pages}{86--94}.
\newblock


\bibitem[Wang et~al\mbox{.}(2019a)]%
        {KGAT}
\bibfield{author}{\bibinfo{person}{Xiang Wang}, \bibinfo{person}{Xiangnan He},
  \bibinfo{person}{Yixin Cao}, \bibinfo{person}{Meng Liu}, {and}
  \bibinfo{person}{Tat{-}Seng Chua}.} \bibinfo{year}{2019}\natexlab{a}.
\newblock \showarticletitle{{KGAT:} Knowledge Graph Attention Network for
  Recommendation}. In \bibinfo{booktitle}{\emph{Proceedings of the 25th {ACM}
  {SIGKDD} International Conference on Knowledge Discovery {\&} Data Mining}}.
  \bibinfo{publisher}{{ACM}}, \bibinfo{pages}{950--958}.
\newblock


\bibitem[Wang et~al\mbox{.}(2019b)]%
        {NGCF}
\bibfield{author}{\bibinfo{person}{Xiang Wang}, \bibinfo{person}{Xiangnan He},
  \bibinfo{person}{Meng Wang}, \bibinfo{person}{Fuli Feng}, {and}
  \bibinfo{person}{Tat{-}Seng Chua}.} \bibinfo{year}{2019}\natexlab{b}.
\newblock \showarticletitle{Neural Graph Collaborative Filtering}. In
  \bibinfo{booktitle}{\emph{Proceedings of the 42nd International {ACM} {SIGIR}
  Conference on Research and Development in Information Retrieval}}.
  \bibinfo{publisher}{{ACM}}, \bibinfo{pages}{165--174}.
\newblock


\bibitem[Wang et~al\mbox{.}(2020)]%
        {DGCF}
\bibfield{author}{\bibinfo{person}{Xiang Wang}, \bibinfo{person}{Hongye Jin},
  \bibinfo{person}{An Zhang}, \bibinfo{person}{Xiangnan He},
  \bibinfo{person}{Tong Xu}, {and} \bibinfo{person}{Tat{-}Seng Chua}.}
  \bibinfo{year}{2020}\natexlab{}.
\newblock \showarticletitle{Disentangled Graph Collaborative Filtering}. In
  \bibinfo{booktitle}{\emph{Proceedings of the 43rd International {ACM} {SIGIR}
  conference on research and development in Information Retrieval}}.
  \bibinfo{publisher}{{ACM}}, \bibinfo{pages}{1001--1010}.
\newblock


\bibitem[Wu et~al\mbox{.}(2021)]%
        {SGL}
\bibfield{author}{\bibinfo{person}{Jiancan Wu}, \bibinfo{person}{Xiang Wang},
  \bibinfo{person}{Fuli Feng}, \bibinfo{person}{Xiangnan He},
  \bibinfo{person}{Liang Chen}, \bibinfo{person}{Jianxun Lian}, {and}
  \bibinfo{person}{Xing Xie}.} \bibinfo{year}{2021}\natexlab{}.
\newblock \showarticletitle{Self-supervised Graph Learning for Recommendation}.
  In \bibinfo{booktitle}{\emph{Proceedings of the 44th International {ACM}
  {SIGIR} Conference on Research and Development in Information Retrieval}}.
  \bibinfo{publisher}{{ACM}}, \bibinfo{pages}{726--735}.
\newblock


\bibitem[Wu et~al\mbox{.}(2022)]%
        {wu2022survey}
\bibfield{author}{\bibinfo{person}{Le Wu}, \bibinfo{person}{Xiangnan He},
  \bibinfo{person}{Xiang Wang}, \bibinfo{person}{Kun Zhang}, {and}
  \bibinfo{person}{Meng Wang}.} \bibinfo{year}{2022}\natexlab{}.
\newblock \showarticletitle{A survey on accuracy-oriented neural
  recommendation: From collaborative filtering to information-rich
  recommendation}.
\newblock \bibinfo{journal}{\emph{IEEE Transactions on Knowledge and Data
  Engineering}} (\bibinfo{year}{2022}), \bibinfo{pages}{In press}.
\newblock


\bibitem[Xia et~al\mbox{.}(2022)]%
        {GNMR}
\bibfield{author}{\bibinfo{person}{Lianghao Xia}, \bibinfo{person}{Chao Huang},
  \bibinfo{person}{Yong Xu}, \bibinfo{person}{Peng Dai},
  \bibinfo{person}{Mengyin Lu}, {and} \bibinfo{person}{Liefeng Bo}.}
  \bibinfo{year}{2022}\natexlab{}.
\newblock \showarticletitle{Multi-Behavior Enhanced Recommendation with
  Cross-Interaction Collaborative Relation Modeling}.
\newblock \bibinfo{journal}{\emph{CoRR}}  \bibinfo{volume}{abs/2201.02307}
  (\bibinfo{year}{2022}).
\newblock


\bibitem[Xia et~al\mbox{.}(2020)]%
        {MATN}
\bibfield{author}{\bibinfo{person}{Lianghao Xia}, \bibinfo{person}{Chao Huang},
  \bibinfo{person}{Yong Xu}, \bibinfo{person}{Peng Dai}, \bibinfo{person}{Bo
  Zhang}, {and} \bibinfo{person}{Liefeng Bo}.} \bibinfo{year}{2020}\natexlab{}.
\newblock \showarticletitle{Multiplex Behavioral Relation Learning for
  Recommendation via Memory Augmented Transformer Network}. In
  \bibinfo{booktitle}{\emph{Proceedings of the 43rd International {ACM} {SIGIR}
  conference on research and development in Information Retrieval}}.
  \bibinfo{publisher}{{ACM}}, \bibinfo{pages}{2397--2406}.
\newblock


\bibitem[Xia et~al\mbox{.}(2021a)]%
        {KHGT}
\bibfield{author}{\bibinfo{person}{Lianghao Xia}, \bibinfo{person}{Chao Huang},
  \bibinfo{person}{Yong Xu}, \bibinfo{person}{Peng Dai}, \bibinfo{person}{Xiyue
  Zhang}, \bibinfo{person}{Hongsheng Yang}, \bibinfo{person}{Jian Pei}, {and}
  \bibinfo{person}{Liefeng Bo}.} \bibinfo{year}{2021}\natexlab{a}.
\newblock \showarticletitle{Knowledge-Enhanced Hierarchical Graph Transformer
  Network for Multi-Behavior Recommendation}. In
  \bibinfo{booktitle}{\emph{Proceedings of the Thirty-Fifth {AAAI} Conference
  on Artificial Intelligence}}. \bibinfo{publisher}{{AAAI} Press},
  \bibinfo{pages}{4486--4493}.
\newblock


\bibitem[Xia et~al\mbox{.}(2021b)]%
        {MBGMN}
\bibfield{author}{\bibinfo{person}{Lianghao Xia}, \bibinfo{person}{Yong Xu},
  \bibinfo{person}{Chao Huang}, \bibinfo{person}{Peng Dai}, {and}
  \bibinfo{person}{Liefeng Bo}.} \bibinfo{year}{2021}\natexlab{b}.
\newblock \showarticletitle{Graph Meta Network for Multi-Behavior
  Recommendation}. In \bibinfo{booktitle}{\emph{Proceedings of the 44th {ACM}
  {SIGIR} International Conference on Research and Development in Information
  Retrieval}}. \bibinfo{publisher}{{ACM}}, \bibinfo{pages}{757--766}.
\newblock


\bibitem[Yan et~al\mbox{.}(2022)]%
        {CRGCN}
\bibfield{author}{\bibinfo{person}{Mingshi Yan}, \bibinfo{person}{Zhiyong
  Cheng}, \bibinfo{person}{Chen Gao}, \bibinfo{person}{Jing Sun},
  \bibinfo{person}{Fan Liu}, \bibinfo{person}{Fuming Sun}, {and}
  \bibinfo{person}{Haojie Li}.} \bibinfo{year}{2022}\natexlab{}.
\newblock \showarticletitle{Cascading Residual Graph Convolutional Network for
  Multi-Behavior Recommendation}.
\newblock \bibinfo{journal}{\emph{CoRR}}  \bibinfo{volume}{abs/2205.13128}
  (\bibinfo{year}{2022}).
\newblock


\bibitem[Yang et~al\mbox{.}(2021)]%
        {HMGGR}
\bibfield{author}{\bibinfo{person}{Haoran Yang}, \bibinfo{person}{Hongxu Chen},
  \bibinfo{person}{Lin Li}, \bibinfo{person}{Philip~S. Yu}, {and}
  \bibinfo{person}{Guandong Xu}.} \bibinfo{year}{2021}\natexlab{}.
\newblock \showarticletitle{Hyper Meta-Path Contrastive Learning for
  Multi-Behavior Recommendation}. In \bibinfo{booktitle}{\emph{Proceedings of
  the 21th {IEEE} International Conference on Data Mining}}.
  \bibinfo{publisher}{{IEEE}}, \bibinfo{pages}{787--796}.
\newblock


\bibitem[Yu et~al\mbox{.}(2022)]%
        {SimGCL}
\bibfield{author}{\bibinfo{person}{Junliang Yu}, \bibinfo{person}{Hongzhi Yin},
  \bibinfo{person}{Xin Xia}, \bibinfo{person}{Tong Chen},
  \bibinfo{person}{Lizhen Cui}, {and} \bibinfo{person}{Quoc Viet~Hung Nguyen}.}
  \bibinfo{year}{2022}\natexlab{}.
\newblock \showarticletitle{Are Graph Augmentations Necessary?: Simple Graph
  Contrastive Learning for Recommendation}. In
  \bibinfo{booktitle}{\emph{Proceedings of the 45th International {ACM} {SIGIR}
  Conference on Research and Development in Information Retrieval}}.
  \bibinfo{publisher}{{ACM}}, \bibinfo{pages}{1294--1303}.
\newblock


\bibitem[Zhang et~al\mbox{.}(2019)]%
        {DLsuvey}
\bibfield{author}{\bibinfo{person}{Shuai Zhang}, \bibinfo{person}{Lina Yao},
  \bibinfo{person}{Aixin Sun}, {and} \bibinfo{person}{Yi Tay}.}
  \bibinfo{year}{2019}\natexlab{}.
\newblock \showarticletitle{Deep Learning Based Recommender System: {A} Survey
  and New Perspectives}.
\newblock \bibinfo{journal}{\emph{{ACM} Comput. Surv.}} \bibinfo{volume}{52},
  \bibinfo{number}{1} (\bibinfo{year}{2019}), \bibinfo{pages}{5:1--5:38}.
\newblock


\bibitem[Zhang et~al\mbox{.}(2020)]%
        {MGNN}
\bibfield{author}{\bibinfo{person}{Weifeng Zhang}, \bibinfo{person}{Jingwen
  Mao}, \bibinfo{person}{Yi Cao}, {and} \bibinfo{person}{Congfu Xu}.}
  \bibinfo{year}{2020}\natexlab{}.
\newblock \showarticletitle{Multiplex Graph Neural Networks for Multi-behavior
  Recommendation}. In \bibinfo{booktitle}{\emph{Proceedings of the 29th {ACM}
  International Conference on Information and Knowledge Management}}.
  \bibinfo{publisher}{{ACM}}, \bibinfo{pages}{2313--2316}.
\newblock


\bibitem[Zhao et~al\mbox{.}(2015)]%
        {zhao2015www}
\bibfield{author}{\bibinfo{person}{Zhe Zhao}, \bibinfo{person}{Zhiyuan Cheng},
  \bibinfo{person}{Lichan Hong}, {and} \bibinfo{person}{Ed~Huai{-}hsin Chi}.}
  \bibinfo{year}{2015}\natexlab{}.
\newblock \showarticletitle{Improving User Topic Interest Profiles by Behavior
  Factorization}. In \bibinfo{booktitle}{\emph{Proceedings of the 24th {WWW}
  International Conference on World Wide Web}}. \bibinfo{publisher}{{ACM}},
  \bibinfo{pages}{1406--1416}.
\newblock


\end{thebibliography}

\end{document}